\documentclass[12pt, centerh1]{article}
\usepackage{natbib}
\usepackage[margin=1in]{geometry}
\usepackage{amsmath,natbib,multirow}
\usepackage{amssymb}
\usepackage{amsfonts,mathrsfs,moreverb,lipsum}
\usepackage{graphicx,colonequals}
\usepackage{color}
\usepackage{caption}
\usepackage{mathtools}
\usepackage{pbox}
\usepackage{multirow}
\usepackage{subfigure}
\usepackage{cleveref}
\usepackage{authblk}
\usepackage[shortlabels]{enumitem}
\usepackage{xcolor}

\Crefname{figure}{Figure}{Figures}

\newcommand{\btheta}{{\boldsymbol{\theta}}}
\newcommand{\bvtheta}{{\boldsymbol{\vartheta}}}

\def\bz{\boldsymbol{z}}

\newcommand{\bphi}{{\boldsymbol{\phi}}}

\newcommand{\vecx}{\mathbf{x}}

\newcommand{\vecX}{\mathbf{X}}

\newcommand{\vecV}{\mathbf{V}}

\newcommand{\vecZ}{\mathbf{Z}}

\newcommand{\matsig}{\mathbf\Sigma}
\newcommand{\matPsi}{\mathbf\Psi}

\newcommand{\vecpi}{\mbox{\boldmath$\pi$}}

\newcommand{\vecR}{\mathbf{R}}

\newcommand{\tr}{\,\mbox{tr}}

\newcommand{\vecA}{\mathbf{A}}

\newcommand{\vecB}{\mathbf{B}}

\newcommand{\vecY}{\mathbf{Y}}
\newcommand{\vecS}{\mathbf{S}}
\newcommand{\vecP}{\mathbf{P}}

\newcommand{\matm}{\mathbf{M}}

\newcommand{\fX}{\mathscr{X}}
\newcommand{\fV}{\mathscr{V}}
\newcommand{\fU}{\mathscr{U}}

\newcommand{\fY}{\mathscr{Y}}

\newcommand{\real}{I\hspace{-1.4mm}R}

\allowdisplaybreaks

\title{Model-based clustering via skewed matrix-variate cluster-weighted models}
\author[1]{Michael P.B. Gallaugher$^*$\thanks{$^*$Corresponding author. Email: Michael\textunderscore Gallaugher@baylor.edu}}
\author[2]{Salvatore D. Tomarchio}
\author[3]{Paul D. McNicholas}
\author[2]{Antonio Punzo}
\affil[1]{Department of Statistical Science, Baylor University, Waco, Texas, USA}
\affil[2]{Department of Economics and Business, University of Catania, Catania, Italy}
\affil[3]{Department of Mathematics and Statistics, McMaster University, Ontario, Canada}

\date{}
\begin{document}
\maketitle{}

\begin{abstract}
Cluster-weighted models (CWMs) extend finite mixtures of regressions (FMRs) in order to allow the distribution of covariates to contribute to the clustering process.
In a matrix-variate framework, the matrix-variate normal CWM has been recently introduced.
However, problems may be encountered when data exhibit skewness or other deviations from normality in the responses, covariates or both. 
Thus, we introduce a family of 24 matrix-variate CWMs which are obtained by allowing both the responses and covariates to be modelled by using one of four existing skewed matrix-variate distributions or the matrix-variate normal distribution. 
Endowed with a greater flexibility, our matrix-variate CWMs are able to handle this kind of data in a more suitable manner.
As a by-product, the four skewed matrix-variate FMRs are also introduced.
Maximum likelihood parameter estimates are derived using an expectation-conditional maximization algorithm.
Parameter recovery, classification assessment, and the capability of the Bayesian information criterion to detect the underlying groups are investigated using simulated data.
Lastly, our matrix-variate CWMs, along with the matrix-variate normal CWM and matrix-variate FMRs, are applied to two real datasets for illustrative purposes.

\end{abstract}

\section{Introduction}
\label{sec:1}

The importance of finite mixture models in statistical data analyses is highlighted by the high volume of articles about mixture applications in the statistical and general scientific literature.
Endowed with great flexibility, the finite mixture model is a convenient statistical tool for the modelling of a wide range of phenomena characterized by unobserved heterogeneity,  and is a very common method for clustering and classification.
Because of its mathematical tractability, the most widespread mixture model makes use of the multivariate normal distribution for the mixture components. 
However, problems may be encountered when data exhibits skewness, atypical observations, or other deviations from normality.
To cope with this issue, an abundance of examples that consider non-normal multivariate component densities have been proposed in the model-based clustering literature (see, e.g. \citealp{peel00,karlis09,lin10,andrews12,dang15,punzo2016p,bagnato2017,tortora19,murray20}).

In addition to the multivariate approaches mentioned above, in the most recent years there has been an increased interest in the area of clustering and classification for matrix-variate data.
This data structure can occur in several and different application domains, such as multivariate longitudinal data, multivariate spatio-temporal data, multivariate repeated measures and multivariate time-series.
In all these cases we have $p$ variables measured in $r$ different occasions on $N$ observations, so that the data can be arranged in a three-way array structure having the following three dimensions: variables (rows), occasions (columns) and observations (layers). 
Examples of contributions to the matrix-variate mixture models literature are \citet{viroli11,viroli2011b,gallaugher18a,gallaugher20,melnykov2018model,melnykov2019studying,sarkar2020parsimonious,tomarchio2020,tomarchio2021b}.

One deficiency of the contributions listed above is that they all fail to take into account possible functional relationships between the variables.
Specifically, important insight can be gained if the variables can be split into response and covariate variables as well as by accounting for a linear relationship among them.
Such a requirement is the basis for the introduction in the literature of finite mixture of regression (FMR) models (see \citealp{desarbo88,fruhwirth06} for examples), that in a matrix-variate setting have been recently presented by \citet{melnykov2019studying}.
The FMR models are also termed as \textit{fixed} covariates approaches, given that they do not explicitly use the distribution of the covariates for clustering.
Put another way, the assignment of data points to clusters does not directly utilize information from the distribution of the covariates.
In contrast, finite mixtures of regression models with \textit{random} covariates \citep{gershenfeld97,gershenfeld1999}, also known as cluster-weighted models (CWMs), allow for assignment dependence: for each mixture component, CWMs decompose the joint distribution of responses and covariates into the product between the conditional distribution of the responses and the marginal distribution of the covariates.
Therefore, the distribution of the covariates affect the assignment of the data points to the clusters.

Over the years, several CWMs have been proposed in the univariate and multivariate model-based clustering literature (see, e.g. \citealp{Ingra12,Ingra12b,Ingra15,subedi13,subedi15,punzo15b,di2020,pocuca20}).
In a matrix-variate framework, \citet{tomarchio2021} recently introduced the matrix-variate normal CWM, where the matrix-variate normal (MVN) distribution is used for modeling both the conditional distribution of the responses and the marginal distribution of the covariates.
As mentioned above, issues may be faced in the presence of skewed data.
For this reason, in this paper we extend this branch of literature by proposing the use of skewed matrix-variate distributions for the conditional distribution of the responses, the marginal distribution of the covariates, or both.
The following skewed matrix-variate distributions recently introduced by \citet{gallaugher17,gallaugher19a} are considered: the matrix-variate skew-t (MVST), the matrix-variate generalized hyperbolic (MVGH), the matrix-variate variance gamma (MVVG), and the matrix-variate normal inverse Gaussian (MVNIG).
By also considering the MVN distribution, we introduce a family of 24 new matrix-variate CWMs that are flexible enough to cope with scenarios where both the responses and the covariates are skewed, or in which one of the two sets of variables is normally distributed and the other is skewed.
Notice that, as a by-product the four skewed matrix-variate FRMs are also introduced.

The remainder of the paper is organized as follows.
In Section~\ref{sec:2}, some preliminary aspects are described.
Section~\ref{sec:3} presents the family of 24 matrix-variate CWMs along with the expectation-conditional maximization (ECM) algorithm \citep{meng93} for parameter estimation.
Section~\ref{sec:sim} discusses two analyses on simulated data, in which the parameter recovery, the classification performances and the capability of the Bayesian information criterion (BIC; \citealp{schwarz78}) to detect the underlying group structure are investigated.
Section~\ref{sec:real} applies our matrix-variate CWMs, along with the matrix-variate normal CWM and matrix-variate FMRs, to two real datasets for illustrative purposes.
Finally, some conclusions and ideas for future developments are drawn in Section~\ref{sec:end}.

\section{Background}
\label{sec:2}

In this section, we present some background concepts used in the development of our matrix-variate CWMs.
In detail, Section~\ref{subsec:2} recalls the generalized inverse Gaussian (GIG) distribution, while Section~\ref{subsec:3} presents the four skewed matrix-variate distributions used in this manuscript.

\subsection{Generalized Inverse Gaussian Distribution} 
\label{subsec:2}

The GIG distribution can be parameterized in several ways \citep{jorgensen82}.
Herein, we will use two different parameterizations.
A random variable $W$ has a GIG distribution with parameters $a>0$, $b>0$, and $\lambda\in \real$, denoted herein by $\mathcal{GIG}(a,b,\lambda)$, if its pdf can be written as
$$
h\left(w;a, b, \lambda\right) = \left(\frac{a}{b}\right)^{\frac{\lambda}{2}}\frac{w^{\lambda-1}}{2K_{\lambda}(\sqrt{ab})}\exp\left[-\frac{1}{2}\left(aw+\frac{b}{w}\right)\right],
$$
where $K_{\lambda}(u)$ is the modified Bessel function of the third kind with index $\lambda$. 

Expectations of some functions of a GIG random variable have a mathematically and computationally tractable form, and will be useful for parameter estimation in Section~\ref{subsec:par_est}. 
Some of such expectations are 
\begin{equation}
\mathbb{E}\left(W\right)=\sqrt{\frac{b}{a}}\frac{K_{\lambda+1}(\sqrt{ab})}{K_{\lambda}(\sqrt{ab})},
\label{eq:ai}
\end{equation}
\begin{equation}
\mathbb{E}\left(\frac{1}{W}\right)=\sqrt{\frac{a}{b}}\frac{K_{\lambda+1}(\sqrt{ab})}{K_{\lambda}(\sqrt{ab})}-\frac{2\lambda}{b},
\label{eq:bi}
\end{equation}
\begin{equation}
\mathbb{E}\left(\log W\right)=\log\left(\sqrt{\frac{b}{a}}\right)+\frac{1}{K_{\lambda}(\sqrt{ab})}\frac{\partial}{\partial \lambda}K_{\lambda}(\sqrt{ab}).
\label{eq:ci}
\end{equation}

An alternative parameterization was used by \cite{browne15} to derive the generalized hyperbolic distribution, and subsequently used by \cite{gallaugher19a} in the matrix-variate case. 
This alternative parameterization is given by
\begin{equation}
h\left(w;\omega,\eta,\lambda\right) = \left(\frac{w}{\eta}\right)^{\lambda-1}\frac{1}{2\eta K_{\lambda}(\omega)}\exp\left[-\frac{\omega}{2}\left(\frac{w}{\eta}+\frac{\eta}{w}\right)\right],
\label{eq:I}
\end{equation}
where $\omega=\sqrt{ab}$, $\eta=\sqrt{b/a}$, and $\lambda$ are concentration, scale, and index parameters, respectively. 
For notational purposes, we will denote the GIG distribution parameterized as in~\eqref{eq:I} by $\mathcal{I}\left(\omega,\eta,\lambda\right)$.

\subsection{Skewed Matrix-Variate Distributions}
\label{subsec:3}
Like in the multivariate and univariate cases, the matrix variate normal is the most well-known matrix variate distribution. A random $p\times r$ matrix $\fX$ follows a matrix variate normal distribution if its probability density function can be written
$$
f(\vecX)=\frac{1}{(2\pi)^{np/2}|\matsig|^{p/2}|\matPsi^{n/2}|}\exp\left\{-\frac{1}{2}\tr(\matsig^{-1}(\vecX-\matm)\matPsi^{-1}(\vecX-\matm)'\right\},
$$
where $\matm$ is the $p\times r$ mean matrix and $\matsig$ and $\matPsi$ are $p\times p$ and $r\times r$ scale matrices, respectively.

A convenient way to obtain skewed matrix-variate distributions is by means of the matrix-variate variance-mean mixture model.
Specifically, this model assumes that a $p\times r$ random matrix $\fV$ can be written as
\begin{equation}
\fV=\matm+W\vecA+\sqrt{W}\fU,
\label{eq:mean-var}
\end{equation}
where $\matm$ is a $p\times r$ location matrix, $\vecA$ is a $p\times r$ skewness matrix, $W$ is a positive random variable and $\fU\sim \mathcal{N}_{p\times r}({\bf 0,\matsig,\matPsi})$ denotes a matrix-variate normal distribution with mean matrix $\matm$, row covariance matrix $\matsig$ and column covariance matrix $\matPsi$.
By changing the distribution of $W$, it is possible to obtain different skewed matrix-variate distributions.
Herein, we consider the following four skewed matrix-variate distributions recently introduced in the literature by \cite{gallaugher17,gallaugher19a}:
\begin{itemize}
\item The MVST distribution, denoted by $\mathcal{MVST}_{p\times r}(\matm,\vecA,\matsig,\matPsi,\nu)$, with pdf 
\begin{align*}
f_{\text{\tiny MVST}}(\vecV;\matm,\vecA,\matsig,\matPsi,\nu)=&\frac{2\left(\frac{\nu}{2}\right)^{\frac{\nu}{2}}\exp\left\{\tr(\matsig^{-1}(\vecV-\matm)\matPsi^{-1}\vecA') \right\} }{(2\pi)^{\frac{pr}{2}}| \matsig |^{\frac{r}{2}} |\matPsi |^{\frac{p}{2}}\Gamma(\frac{\nu}{2})}  \left(\frac{\delta(\vecV;\matm,\matsig,\matPsi)+\nu}{\rho(\vecA,\matsig,\matPsi)}\right)^{-\frac{\nu+pr}{4}} \\ & \times K_{-\frac{\nu+pr}{2}}\left(\sqrt{\left[\rho(\vecA,\matsig,\matPsi)\right]\left[\delta(\vecV;\matm,\matsig,\matPsi)+\nu\right]}\right),
\end{align*}
where 
$
\delta(\vecV;\matm,\matsig,\matPsi)=\tr(\matsig^{-1}(\vecV-\matm)\matPsi^{-1}(\vecV-\matm)')$, $\rho(\vecA,\matsig,\matPsi)=\tr(\matsig^{-1}\vecA\matPsi^{-1}\vecA')
$
and $\nu>0$;
\item The MVGH distribution, denoted by $\mathcal{MVGH}_{p\times r}(\matm,\vecA,\matsig,\matPsi,\lambda,\omega)$, having pdf
\begin{align*}
f_{\text{\tiny MVGH}}(\vecV;\matm,\vecA,\matsig,\matPsi,\lambda,\omega)=&\frac{\exp\left\{\tr(\matsig^{-1}(\vecV-\matm)\matPsi^{-1}\vecA') \right\}}{(2\pi)^{\frac{pr}{2}}| \matsig |^{\frac{r}{2}} |\matPsi |^{\frac{p}{2}}K_{\lambda}(\omega)}  \left(\frac{\delta(\vecV;\matm,\matsig,\matPsi)+\omega}{\rho(\vecA,\matsig,\matPsi)+\omega}\right)^{\frac{\left(\lambda-\frac{pr}{2}\right)}{2}} \\ & \times
 K_{\left(\lambda-{pr}/{2}\right)}\left(\sqrt{\left[\rho(\vecA,\matsig,\matPsi)+\omega\right]\left[\delta(\vecV;\matm,\matsig,\matPsi)+\omega\right]}\right)
\end{align*}
with $\lambda\in\real$ and $\omega>0$;
\item The MVVG distribution, denoted by $\mathcal{MVVG}_{p\times r}(\matm,\vecA,\matsig,\matPsi,\gamma)$, with pdf 
\begin{align*}
f_{\text{\tiny MVVG}}(\vecV,\matm,\vecA,\matsig,\matPsi,\gamma)=&\frac{2\gamma^{\gamma}\exp\left\{\tr(\matsig^{-1}(\vecV-\matm)\matPsi^{-1}\vecA') \right\}}{(2\pi)^{\frac{pr}{2}}| \matsig |^{\frac{r}{2}} |\matPsi |^{\frac{p}{2}}\Gamma(\gamma)}  \left(\frac{\delta(\vecV;\matm,\matsig,\matPsi)}{\rho(\vecA,\matsig,\matPsi)+2\gamma}\right)^{\frac{\left(\gamma-{pr}/{2}\right)}{2}} \\
&\times  K_{\left(\gamma-\frac{pr}{2}\right)}\left(\sqrt{\left[\rho(\vecA,\matsig,\matPsi)+2\gamma\right]\left[\delta(\vecV;\matm,\matsig,\matPsi)\right]}\right),
\end{align*}
where $\gamma>0$; and
\item The MVNIG distribution, denoted by $\mathcal{MVNIG}_{p\times r}(\matm,\vecA,\matsig,\matPsi,\kappa)$, having pdf 
\begin{align*}
f_{\text{\tiny MVNIG}}(\vecV;\matm,\vecA,\matsig,\matPsi,\kappa)&=\frac{2\exp\left\{\tr(\matsig^{-1}(\vecV-\matm)\matPsi^{-1}\vecA')+\kappa\right\}
}{(2\pi)^{\frac{pr+1}{2}}| \matsig |^{\frac{r}{2}} |\matPsi |^{\frac{p}{2}}}\left(\frac{\delta(\vecV;\matm,\matsig,\matPsi)+1}{\rho(\vecA,\matsig,\matPsi)+\kappa^2}\right)^{-{\left(1+pr\right)}/{4}}\\
&\times K_{-{(1+pr)}/{2}}\left(\sqrt{\left[\rho(\vecA,\matsig,\matPsi)+\kappa^2\right]\left[\delta(\vecV;\matm,\matsig,\matPsi)+1\right]}\right),
\end{align*}
with $\kappa>0$. 
\end{itemize}

%
%
%
%
%
%
%

\section{Methodology}
\label{sec:3}

\subsection{CWMs using Matrix-Variate Skewed Distributions}
\label{sec:3.1}

Let $\fY$ be a continuous random matrix of dimension $p \times r$, containing $p$ responses measured over $r$ occasions. 
Consider a $\fX$ continuous random matrix of dimension $q \times r$, containing $q$ covariates evaluated over $r$ occasions. 
Furthermore, assume there exist $G$ subgroups in the data. 
Then, in a matrix-variate CWM framework, the joint pdf of $\fY$ and $\fX$ can be written as
\begin{equation}
p(\vecY,\vecX;\bvtheta)=\sum_{g=1}^G \pi_g f(\vecY|\vecX;\bvtheta_{\vecY|g}) f(\vecX;\bvtheta_{\vecX|g}),
\label{eq:cwmg}
\end{equation}
where $\pi_g>0$ are the mixing proportions, with $\sum_{g=1}^G\pi_g=1$, $f(\vecY|\vecX;\bvtheta_{\vecY|g})$ is the conditional distribution of the responses, $f(\vecX;\bvtheta_{\vecX|g})$ is the distribution of the covariates, and $\bvtheta=\left\{\pi_g,\bvtheta_{\vecY|g},\bvtheta_{\vecX|g};g=1,\ldots,G\right\}$ contains all the parameters of the model.

The dependence of $\fY$ on $\fX=\vecX$ in the $g$th mixture component is typically accounted for by allowing the mean or, more generally, a location parameter in $\bvtheta_{\vecY|g}$, to depend on $\vecX$ via some linear functional relationship.
The possibility to specify different models for either $f(\vecY|\vecX;\bvtheta_{\vecY|g})$ or $f(\vecX;\bvtheta_{\vecX|g})$ makes the matrix-variate CWM a very flexible modelling approach.
As mentioned in Section~\ref{sec:1}, \citet{tomarchio2021} assume for the $g$th mixture component a MVN distribution for $\fX$, say $\fX\sim \mathcal{N}_{q\times r}({\matm_{\vecX|g},\matsig_{\vecX|g},\matPsi_{\vecX|g}})$, and a MVN distribution for $\fY|\vecX$, say $\fY|\vecX\sim \mathcal{N}_{p\times r}({\matm_{\vecY|g}\left(\vecX;\vecB_g\right),\matsig_{\vecY|g},\matPsi_{\vecY|g}})$.
Furthermore, it is assumed a linear relation $\matm_{\vecY|g}\left(\vecX;\vecB_g\right)=\vecB_g\vecX^*$, where $\vecB_g$ is a $p \times (1 + q)$ matrix of regression coefficients and 
$\vecX^*$ is the $ (1 + q) \times r$ matrix containing the intercept and the $q$ covariates.

In this paper, we assume that $f(\vecY|\vecX;\bvtheta_{\vecY|g})$ and $f(\vecX;\bvtheta_{\vecX|g})$ in~\eqref{eq:cwmg} can be any of the four skewed matrix-variate distributions discussed in Section~\ref{subsec:3} or the MVN distribution.
Being members of the family of distributions defined by \eqref{eq:mean-var}, in the $g$th mixture component we have 
\begin{equation}
\begin{split}
\fY|\vecX & = \vecB_g\vecX^*+W_{\vecY|g}\vecA_{\vecY|g}+\sqrt{W_{\vecY|g}}\fU_{\vecY|g}, \\
\fX & = \matm_{\vecX|g}+W_{\vecX|g}\vecA_{\vecX|g}+\sqrt{W_{\vecX|g}}\fU_{\vecX|g}.
\end{split}
\label{eq:vmm2}
\end{equation}
Considering that $f(\vecY|\vecX;\bvtheta_{\vecY|g})$ and $f(\vecX;\bvtheta_{\vecX|g})$ are not required to be the same, we obtain a family of 25 matrix-variate CWMs, 24 of which are herein introduced.
For the purposes of notation, each model will be labeled by separating with a dash the acronyms used for $f(\vecX;\bvtheta_{\vecX|g})$ and $f(\vecY|\vecX;\bvtheta_{\vecY|g})$, respectively.
For example, if we consider a matrix-variate CWM having a MVST distribution for $\fX$ and a MVVG distribution for $\fY|\vecX$, it will be referred to as MVST-MVVG CWM.

\subsection{Parameter Estimation}
\label{subsec:par_est}

Parameter estimation is carried out via the ECM algorithm, which differs from the expectation-maximization (EM) algorithm \citep{dempster77} because the M-step is replaced by a sequence of simpler and computationally convenient CM-steps. 
The EM algorithm cannot be directly implemented because there is no closed form solution for the covariance matrices of matrix-variate distributions, i.e., one of the two depends on the value of the other at the previous iteration.

Let $\vecS = \left\{\left(\vecY_{i},\vecX_{i}\right)\right\}_{i=1}^{N}$ be a sample of $N$ independent observations from model~\eqref{eq:cwmg}.
Within the formulation of model~\eqref{eq:cwmg}, $\vecS$ is viewed as being incomplete and we have two sources of incompleteness.
The first source arises from the fact that, for each observation, we do not know its component membership; we use an indicator vector $\bz_i=\left(z_{i1},\ldots,z_{iG}\right)$, where $z_{ig}=1$ if observation $i$ is in group $g$, and $z_{ig}=0$ otherwise, to govern this source.
The second source arises if $f(\vecY|\vecX;\bvtheta_{\vecY|g})$ or $f(\vecX;\bvtheta_{\vecX|g})$ are skewed; we need the latent variables $W_{\vecY|g}$ and $W_{\vecX|g}$ introduced in~\eqref{eq:vmm2} to govern this source.
Considering this, we can write the complete-data log likelihood as
\begin{equation}
l(\bvtheta)=l_1(\vecpi)+l_2(\btheta_{\vecX})+l_3(\btheta_{\vecY}),
  \label{eq:compl_llk}
\end{equation}
where $\vecpi=(\pi_1,\ldots,\pi_G)'$, and 
$$
l_1(\vecpi)=\sum_{g=1}^G\sum_{i=1}^Nz_{ig}\log(\pi_g).
$$
If the covariates $\fX$ are assumed to be one of the four skewed matrix-variate distributions, then
\begin{align*}
l_2(\btheta_{\vecX}) 
= &
\sum_{g=1}^G\sum_{i=1}^N z_{ig} \log\left[h(w_{ig\vecX};\bphi_{W_{\vecX}|g})\right] + C_{\vecX}-\frac{1}{2}\sum_{g=1}^G\sum_{i=1}^N z_{ig} \left\{r\log(|\matsig_{\vecX|g}|)+q\log(|\matPsi_{\vecX|g}|)\right.\\
& +\tr\left[(1/w_{ig\vecX})\matsig_{\vecX|g}^{-1}(\vecX_i-\matm_{\vecX|g})\matPsi_{\vecX|g}^{-1}(\vecX_i-\matm_{\vecX|g})'-\matsig_{\vecX|g}^{-1}(\vecX_i-\matm_{\vecX|g})\matPsi_{\vecX|g}^{-1}\vecA_{\vecX|g}'\right.\\&\left.\left.-\matsig_{\vecX|g}^{-1}\vecA_{\vecX|g}\matPsi_{\vecX|g}^{-1}(\vecX_i-\matm_{\vecX|g})'+w_{ig\vecX}\matsig_{\vecX|g}^{-1}\vecA_{\vecX|g}\matPsi_{\vecX|g}^{-1}\vecA_{\vecX|g}'\right]\right\},
\end{align*}
where $h(w_{ig\vecX};\bphi_{W_{\vecX}|g})$ is the appropriate pdf for $W_{ig\vecX}$ presented in Section~\ref{subsec:3}, with parameters denoted as $\bphi_{W_{\vecX}|g}$, while $C_{\vecX}$ is constant with respect to the parameters.
Otherwise, if the covariates $\fX$ are assumed to be normally distributed, then 
$$
l_2(\btheta_{\vecX}) 
=C_{\vecX}-\frac{1}{2}\sum_{g=1}^G\sum_{i=1}^Nz_{ig}[r\log(|\matsig_{\vecX|g}|)+q\log(|\matPsi_{\vecX|g}|)+\tr(\matsig_{\vecX|g}^{-1}(\vecX_i-\matm_{\vecX|g})\matPsi_{\vecX|g}^{-1}(\vecX_i-\matm_{\vecX|g})')].
$$

In a similar way, if the conditional distribution of $\fY|\vecX$ is of the four skewed matrix-variate distributions, then
\begin{align*}
l_3(\btheta_{\vecY}) 
= & 
\sum_{g=1}^G\sum_{i=1}^N \log\left[h(w_{ig\vecY};\bphi_{W_{\vecY}|g})\right]+ C_{\vecY} -\frac{1}{2}\sum_{g=1}^G\sum_{i=1}^N z_{ig} \left\{r\log(|\matsig_{\vecY|g}|)+p\log(|\matPsi_{\vecY|g}|)\right. \\& +\tr\left[(1/w_{ig\vecY})\matsig_{\vecY|g}^{-1}(\vecY_i-{\vecB_g\vecX_i^*})\matPsi_{\vecY|g}^{-1}(\vecY_i-{\vecB_g\vecX_i^*})'-\matsig_{\vecY|g}^{-1}(\vecY_i-{\vecB_g\vecX_i^*})\matPsi_{\vecY|g}^{-1}\vecA_{\vecY|g}'\right.\\&\left.\left.-\matsig_{\vecY|g}^{-1}\vecA_{\vecY|g}\matPsi_{\vecY|g}^{-1}(\vecY_i-{\vecB_g\vecX_i^*})' + w_{ig\vecY}\matsig_{\vecY|g}^{-1}\vecA_{\vecY|g}\matPsi_{\vecY|g}^{-1}\vecA_{\vecY|g}'\right]\right\}.
\end{align*}

Otherwise, if $\fY|\vecX$ is assumed to be normally distributed, then
$$
l_3(\btheta_{\vecY}) 
= C_{\vecY} -\frac{1}{2}\sum_{g=1}^G\sum_{i=1}^N z_{ig}[r\log(|\matsig_{\vecY|g}|)+p\log(|\matPsi_{\vecY|g}|)+\matsig_{\vecY|g}^{-1}(\vecY_i-\vecB_g\vecX_i^*)\matPsi_{\vecY|g}^{-1}(\vecY_i-\vecB_g\vecX_i^*)'].
$$

In the following, by adopting the notation used in \citet{tomarchio2021}, the quantities marked with one dot correspond to the updates at the previous iteration and those marked with two dots represent the updates at the current iteration.

\paragraph{E-Step}
The E-step requires the calculation of the conditional expectation of~\eqref{eq:compl_llk}.
Therefore, we first need to compute
$$
\ddot{z}_{ig}=\frac{\dot{\pi}_g f\left(\vecY_i|\vecX_i;\dot{\btheta}_{\vecY|g}\right)f\left(\vecX_i;\dot{\btheta}_{\vecX|g}\right)}{\displaystyle\sum_{h=1}^G\dot{\pi}_h f\left(\vecY_i|\vecX_i;\dot{\btheta}_{\vecY|h}\right)f\left(\vecX_i;\dot{\btheta}_{\vecX|h}\right)},
$$
that is the posterior probability that the unlabeled observation $\left(\vecX_{i},\vecY_{i}\right)$ belongs to the $g$th component of the CWM.
Then, if the distribution of $\fX$ in component $g$, $g = 1,\ldots, G$, is skewed, the following quantities must be computed
\begin{equation*}
\begin{split}
\ddot{l}_{ig\vecX}&\colonequals\mathbb{E}[W_{ig\vecX}|z_{ig}=1,\vecX_i,\dot{\bphi}_{W_{\vecX}|g}],\\
\ddot{m}_{ig\vecX}&\colonequals\mathbb{E}[1/W_{ig\vecX}|z_{ig}=1,\vecX_i,\dot{\bphi}_{W_{\vecX}|g}],\\
\ddot{n}_{ig\vecX}&\colonequals\mathbb{E}[\log(W_{ig\vecX})|z_{ig}=1,\vecX_i,\dot{\bphi}_{W_{\vecX}|g}].\\
\end{split}
\end{equation*} 
Furthermore, if the distribution of $\fY|\vecX$ is skewed, then the following values are also updated:
\begin{equation*}
\begin{split}
\ddot{l}_{ig\vecY}&\colonequals\mathbb{E}[W_{ig\vecY}|z_{ig}=1,\vecY_i,\vecX_i,\dot{\bphi}_{W_{\vecY}|g}],\\
\ddot{m}_{ig\vecY}&\colonequals\mathbb{E}[1/W_{ig\vecY}|z_{ig}=1,\vecY_i,\vecX_i,\dot{\bphi}_{W_{\vecY}|g}],\\
\ddot{n}_{ig\vecY}&\colonequals\mathbb{E}[\log(W_{ig\vecY})|z_{ig}=1,\vecY_i,\vecX_i,\dot{\bphi}_{W_{\vecY}|g}].\\
\end{split}
\end{equation*}
By inserting a superscript to $W_{ig\vecX}$ and $W_{ig\vecY}$ with the label associated to the considered skewed matrix-variate distribution, we have that
\begin{align*}
W^{\text{MVST}}_{ig\vecX}~|~\vecX_i, z_{ig}=1&\sim \mathcal{GIG}\left(\rho(\vecA_{\vecX|g},\matsig_{\vecX|g}),\delta(\vecX_i;\matm_{\vecX|g},\matsig_{\vecX|g})+\nu_{\vecX|g},-(\nu_{\vecX|g}+qr)/2\right) \\ 
W^{\text{MVGH}}_{ig\vecX}~|~\vecX_i, z_{ig}=1&\sim \mathcal{GIG}\left(\rho(\vecA_{\vecX|g},\matsig_{\vecX|g})+{\omega_{X}}_{g},\delta(\vecX_i;\matm_{\vecX|g},\matsig_{\vecX|g})+\omega_{\vecX|g},\lambda_{\vecX|g}-{qr}/{2}\right) \\
W^{\text{MVVG}}_{ig\vecX}~|~\vecX_i, z_{ig}=1&\sim \mathcal{GIG}\left(\rho(\vecA_{\vecX|g},\matsig_{\vecX|g})+2\gamma_{\vecX|g},\delta(\vecX_i;\matm_{\vecX|g},\matsig_{\vecX|g}),{\gamma_{X}}_g-{qr}/{2}\right) \\
W^{\text{MVNIG}}_{ig\vecX}~|~\vecX_i, z_{ig}=1&\sim \mathcal{GIG}\left(\rho(\vecA_{\vecX|g},\matsig_{\vecX|g})+{\kappa^2_{X}}_g,\delta(\vecX_i;\matm_{\vecX|g},\matsig_{\vecX|g})+1,-{(1+qr)}/{2}\right) 
\end{align*}
and 
\begin{align*}
W^{\text{MVST}}_{ig\vecY}~|~\vecY_i, \vecX_i, z_{ig}=1&\sim \mathcal{GIG}\left(\rho(\vecA_{\vecY|g},\matsig_{\vecY|g}),\delta(\vecY_i;\vecB_g\vecX_i^*,\matsig_{\vecY|g})+\nu_{\vecY|g},-(\nu_{\vecY|g}+pr)/2\right) \\ 
W^{\text{MVGH}}_{ig\vecX}~|~\vecY_i, \vecX_i, z_{ig}=1&\sim \mathcal{GIG}\left(\rho(\vecA_{\vecY|g},\matsig_{\vecY|g})+{\omega_{Y}}_{g},\delta(\vecY_i;\vecB_g\vecX_i^*,\matsig_{\vecY|g})+\omega_{\vecY|g},\lambda_{\vecY|g}-{pr}/{2}\right) \\
W^{\text{MVVG}}_{ig\vecY}~|~\vecY_i, \vecX_i, z_{ig}=1&\sim \mathcal{GIG}\left(\rho(\vecA_{\vecY|g},\matsig_{\vecY|g})+2\gamma_{\vecY|g},\delta(\vecY_i;\vecB_g\vecX_i^*,\matsig_{\vecY|g}),{\gamma_{X}}_g-{pr}/{2}\right)  \\
W^{\text{MVNIG}}_{ig\vecY}~|~\vecY_i, \vecX_i, z_{ig}=1&\sim \mathcal{GIG}\left(\rho(\vecA_{\vecY|g},\matsig_{\vecY|g})+{\kappa^2_{X}}_g,\delta(\vecY_i;\vecB_g\vecX_i^*,\matsig_{\vecY|g})+1,-{(1+pr)}/{2}\right) 
\end{align*}
Therefore, all of the required expectations can be calculated using \eqref{eq:ai}--\eqref{eq:ci}.

\paragraph{First CM-Step}

In the first CM step, we maximize the expectation of the complete-data log-likelihood with respect to $\bvtheta_1=\left\{\pi_g,\matm_{\vecX|g},\vecB_g,\vecA_{\vecX|g},\vecA_{\vecY|g},\matsig_{\vecX|g},\matsig_{\vecY|g}\right\}_{g=1}^{G}$, fixing $\bvtheta_2=\left\{\matPsi_{\vecX|g},\matPsi_{\vecY|g}\right\}$ at $\dot\bvtheta_2$.
Notice that $\vecA_{\vecX|g}$ and $\vecA_{\vecY|g}$ are updated only in the case of skewed matrix-variate distributions. 
The update for $\pi_g$ is
\begin{equation*}
\ddot{\pi}_g = \frac{1}{N}\sum_{i=1}^N\ddot{z}_{ig}.
\end{equation*}
The parameters related to the distribution of $\fX$ are updated as follows. 
If, in component~$g$, $g = 1,\ldots, G$, $\fX$ follows one of the four skewed matrix-variate distributions, we have the following updates:
\begin{equation*}
\ddot{\matm}_{\vecX|g}=\frac{\sum_{i=1}^N\ddot{z}_{ig}\vecX_i\left(\overline{l}_{\vecX|g} \ddot{m}_{ig\vecX}-1\right)}{\sum_{i=1}^N\ddot{z}_{ig}\overline{l}_{\vecX|g} \ddot{m}_{ig\vecX}-\ddot{T}_g},\qquad
\ddot{\vecA}_{\vecX|g}=\frac{\sum_{i=1}^N\ddot{z}_{ig} \vecX_i\left(\overline{m}_{\vecX|g}-\ddot{m}_{ig\vecX}\right)}{\sum_{i=1}^N\ddot{z}_{ig}\overline{l}_{\vecX|g} \ddot{m}_{ig\vecX}-\ddot{T}_g},
\end{equation*}
\begin{align*}
\ddot{\matsig}_{\vecX|g} & = \frac{1}{r \ddot{T}_g}\sum_{i=1}^N\ddot{z}_{ig}\left[\ddot{m}_{ig\vecX} (\vecX_i-\ddot{\matm}_{\vecX|g})\dot{\matPsi}_{\vecX|g}^{-1}(\vecX_i-\ddot{\matm}_{\vecX|g})'-(\vecX_i-\ddot{\matm}_{\vecX|g})\dot{\matPsi}_{\vecX|g}^{-1}\ddot{\vecA}_{\vecX|g}'\right.\\ & \left. -\ddot{\vecA}_{\vecX|g}\dot{\matPsi}_{\vecX|g}^{-1}(\vecX_i-\ddot{\matm}_{\vecX|g})'+ \ddot{l}_{ig\vecX}\ddot{\vecA}_{\vecX|g}\dot{\matPsi}_{\vecX|g}^{-1}\ddot{\vecA}_{\vecX|g}'\right],
\end{align*}
where $\ddot{T}_g=\sum_{i=1}^N\ddot{z}_{ig}$, $\overline{l}_{\vecX|g}=(1/\ddot{T}_g)\sum_{i=1}^N\ddot{z}_{ig}\ddot{l}_{ig\vecX}$ and $\overline{m}_{\vecX|g}=(1/\ddot{T}_g)\sum_{i=1}^N\ddot{z}_{ig}\ddot{m}_{ig\vecX}$.
On the other hand, if in component $g$, $g = 1,\ldots, G$, $\fX$ is normally distributed then
\begin{equation*}
\ddot{\matm}_g=\frac{1}{\ddot{T}_g}\sum_{g=1}^G\ddot{z}_{ig}\vecX_i, \qquad 
\ddot{\matsig}_{\vecX|g}=\frac{1}{r\ddot{T}_g}\sum_{g=1}^G\ddot{z}_{ig}(\vecX_i-\ddot{\matm}_{\vecX|g})\dot{\matPsi}_{\vecX|g}^{-1}(\vecx_i-\ddot{\matm}_{\vecX|g})'. 
\end{equation*}

The parameters related to the distribution of $\fY|\vecX$ are updated as follows. 
For the four skewed matrix-variate distributions, we have the following updates 
$$
\ddot{\vecB}_g= \ddot{\vecR}_g \ddot{\vecP}_g^{-1}, \qquad \ddot{\vecA}_{\vecY|g}=\frac{1}{\ddot{T}_g \overline{l}_{\vecY|g}}\left(\sum_{i=1}^N\ddot{z}_{ig}\vecY_i-\ddot{\vecR}_g \ddot{\vecP}_g^{-1}\sum_{i=1}^N\ddot{z}_{ig}\vecX_i^*\right),
$$
\begin{align*}
\ddot{\matsig}_{\vecY|g} & = \frac{1}{r \ddot{T}_g}\sum_{i=1}^N\ddot{z}_{ig}\left[\ddot{m}_{ig\vecY} (\vecY_i-\ddot{\vecB}_g\vecX_i^*)\dot{\matPsi}_{\vecY|g}^{-1}(\vecY_i-\ddot{\vecB}_g\vecX_i^*)'-(\vecY_i-\ddot{\vecB}_g\vecX_i^*)\dot{\matPsi}_{\vecY|g}^{-1}\ddot{\vecA}_{\vecY|g}'\right.\\ & \left. -\ddot{\vecA}_{\vecY|g}\dot{\matPsi}_{\vecY|g}^{-1}(\vecY_i-\ddot{\vecB}_g\vecX_i^*)'+ \ddot{l}_{ig\vecY}\ddot{\vecA}_{\vecY|g}\dot{\matPsi}_{\vecY|g}^{-1}\ddot{\vecA}_{\vecY|g}'\right],
\end{align*}
where 
$$
\ddot{\vecP}_g=\sum_{i=1}^N\ddot{z}_{ig} \ddot{m}_{ig\vecY}\vecX_i^*\dot{\matPsi}_{\vecY|g}^{-1}{\vecX_i^*}'-\frac{1}{\ddot{T}_g\overline{l}_g}\left(\sum_{i=1}^N\ddot{z}_{ig}\vecX_i^*\right)\dot{\matPsi}_{\vecY|g}^{-1}\left(\sum_{i=1}^N\ddot{z}_{ig}{\vecX_i^*}'\right),
$$
$$
\ddot{\vecR}_g=\sum_{i=1}^N\ddot{z}_{ig}\ddot{m}_{ig\vecY}\vecY_i\dot{\matPsi}_{\vecY|g}^{-1}{\vecX_i^*}'-\frac{1}{\ddot{T}_g\overline{l}_g}\left(\sum_{i=1}^N\ddot{z}_{ig}\vecY_i\right)\dot{\matPsi}_{\vecY|g}^{-1}\left(\sum_{i=1}^N\ddot{z}_{ig}{\vecX_i^*}'\right),
$$
and $\overline{l}_{\vecY|g}=(1/\ddot{T}_g)\sum_{i=1}^N\ddot{z}_{ig}\ddot{l}_{ig\vecY}$.

Conversely, if $\fY|\vecX$ is normally distributed then
$$
\ddot{\vecB}_g=\left(\sum_{i=1}^N\ddot{z}_{ig}\vecY_i\dot{\matPsi}_{\vecY|g}^{-1}{\vecX_i^*}\right)\left(\sum_{i=1}^N\ddot{z}_{ig}\vecX_i^*\dot{\matPsi}_{\vecY|g}^{-1}{\vecX_i^*}'\right)^{-1}
$$
and
$$
\ddot{\matsig}_{\vecY|g}=\frac{1}{r\ddot{T}_g}\sum_{g=1}^G\ddot{z}_{ig}(\vecX_i-\ddot{\vecB}_g\vecX_i^*)\dot{\matPsi}_{\vecY|g}^{-1}(\vecx_i-\ddot{\vecB}_g\vecX_i^*)'. 
$$

\paragraph{Second CM-Step}

In the second CM-step, we maximize the expectation of the complete-data log-likelihood with respect to $\bvtheta_2$, keeping fixed $\bvtheta_1$ at $\ddot{\bvtheta}_1$.
Thus, if in component $g$, $g = 1,\ldots, G$, $\fX$ follows one of the four skewed matrix-variate distributions, we have the following update 
\begin{align*}
\ddot{\matPsi}_{\vecX|g} & = \frac{1}{q \ddot{T}_g}\sum_{i=1}^N\ddot{z}_{ig}\left[\ddot{m}_{ig\vecX}(\vecX_i-\ddot{\matm}_{\vecX|g})'\ddot{\matsig}_{\vecX|g}^{-1}(\vecX_i-\ddot{\matm}_{\vecX|g})-(\vecX_i-\ddot{\matm}_{\vecX|g})'\ddot{\matsig}_{\vecX|g}^{-1}\ddot{\vecA}_{\vecX|g}\right.\\&\left.-\ddot{\vecA}_{\vecX|g}'\ddot{\matsig}_{\vecX|g}^{-1}(\vecX_i-\ddot{\matm}_{\vecX|g})+\ddot{l}_{ig\vecX}\ddot{\vecA}_{\vecX|g}'\ddot{\matsig}_{\vecX|g}^{-1}\ddot{\vecA}_{\vecX|g}\right].
\end{align*}
On the contrary, if, in component $g$, $g = 1,\ldots, G$, $\fX$ is assumed normal, then
$$
\ddot{\matPsi}_{\vecX|g}=\frac{1}{q\ddot{T}_g}\sum_{g=1}^G\ddot{z}_{ig}(\vecX_i-\ddot{\matm}_{\vecX|g})'\ddot{\matsig}_{\vecX|g}^{-1}(\vecX_i-\ddot{\matm}_{\vecX|g}).
$$

If, in component $g$, $g = 1,\ldots, G$, $\fY|\vecX$ is one of the four skewed matrix-variate distributions, we have the following update 
\begin{align*}
\ddot{\matPsi}_{\vecY|g} & = \frac{1}{p \ddot{T}_g}\sum_{i=1}^N\ddot{z}_{ig}\left[\ddot{m}_{ig\vecY}(\vecY_i-\ddot{\vecB}_g\vecX_i^*)'\ddot{\matsig}_{\vecY|g}^{-1}(\vecY_i-\ddot{\vecB}_g\vecX_i^*)-(\vecY_i-\ddot{\vecB}_g\vecX_i^*)'\ddot{\matsig}_{\vecY|g}^{-1}\ddot{\vecA}_{\vecY|g}\right.\\&\left.-\ddot{\vecA}_{\vecY|g}'\ddot{\matsig}_{\vecY|g}^{-1}(\vecY_i-\ddot{\vecB}_g\vecX_i^*)+\ddot{l}_{ig\vecY}\ddot{\vecA}_{\vecY|g}'\ddot{\matsig}_{\vecY|g}^{-1}\ddot{\vecA}_{\vecY|g}\right].
\end{align*}

Otherwise, if, in component $g$, $g = 1,\ldots, G$, $\fY|\vecX$ is assumed normal, then
$$
\ddot{\matPsi}_{\vecY|g}=\frac{1}{p\ddot{T}_g}\sum_{i=1}^N\ddot{z}_{ig}(\vecY_i-\ddot{\vecB}_g\vecX_i^*)'\ddot{\matsig}_{\vecY|g}^{-1}(\vecY_i-\ddot{\vecB}_g\vecX_i^*).
$$

\paragraph{Third CM-Step}

In the third CM-step, the additional parameters related to $W$ are now updated both for $\fX$ and $\fY|\vecX$. 
These updates will vary according to the considered skewed matrix-variate distribution, and are reported in \appendixname~\ref{sec:App1} to avoid an excessive length of this section.

\subsection{A note on the ECM initialization}
\label{sec:init}

To start the ECM algorithm, we initialize the $z_{ig}$ in two different ways:
\begin{itemize}
\item In a ``soft'' way, by generating $G$ positive random values from a uniform distribution on [0,1] for each observation, that are subsequently normalized to have
a unitary sum. Being purely random, this procedure is repeated 9 times, and the solution maximizing the observed-data log-likelihood among these runs is considered.
\item In a ``hard'' way, by using the classification produced by the $k$-means algorithm on the vectorized and merged data.
\end{itemize}
The approach providing the largest (observed data) log-likelihood is then selected.

\section{Simulated Data Analyses}
\label{sec:sim}

In this section, the parameter recovery of our algorithm, the classification performance of the matrix-variate CWMs and the capability of the Bayesian information
criterion to detect the underlying group structure, are evaluated.
To assess the parameter recovery, we consider the mean squared error (MSE). 
We use the adjusted Rand index \citep[ARI;][]{{hubert85}} to evaluate the classification performance.
We recall that the ARI can be used to calculate the agreement between the true classification and the one produced by the model. An ARI of 1 indicates perfect agreement between them, whereas the expected value of the ARI under random classification is 0.

\subsection{Parameter Recovery}
\label{sec:pr}

Considering the high number of matrix-variate CWMs introduced in this work, we focus our attention on a subset of four models.
We select the models such that the following different cases are covered:
\begin{enumerate}
\item $f(\vecX;\bvtheta_{\vecX|g})$ and $f(\vecY|\vecX;\bvtheta_{\vecY|g})$ are the same skewed density;
\item $f(\vecX;\bvtheta_{\vecX|g})$ and $f(\vecY|\vecX;\bvtheta_{\vecY|g})$ are different skewed densities,
\item $f(\vecX;\bvtheta_{\vecX|g})$ is skewed and $f(\vecY|\vecX;\bvtheta_{\vecY|g})$ is normal;
\item $f(\vecX;\bvtheta_{\vecX|g})$ is normal and $f(\vecY|\vecX;\bvtheta_{\vecY|g})$ is skewed.
\end{enumerate}
As illustrative examples, we consider (1) the MVVG-MVVG CWM, (2) the MVGH-MVST CWM, (3) the MVNIG-MVN CWM and (4) the MVN-MVGH CWM.
These models are chosen so that all the distributions considered in this manuscript are incorporated in some way.

Because of the high number of parameters involved, we limit our discussion to the recovery of the regression coefficients, as commonly done in the CWM literature (see, e.g.~\citealp{punzo14b,punzo14c,Ingra15,punzo15b}). 
For each model we consider $p=3$, $r=4$, $q=3$ and $G=3$.
Two sample sizes are considered ($N=200$ and $N=500$), as well as two levels of separation that we refer to as ``close'' and ``far''.
With respect to the levels of separation, we assume that the groups have the same parameters with the exclusion of $\matm_{\vecX|g}$.
In this way, we are able to reach different levels of overlap by simply controlling the location parameter for the covariates.
Therefore, each pair ($N$, overlap) leads to four scenarios for each of the aforementioned matrix-variate CWMs.
The parameters used to generate the data in every scenario, and that are the same for the four matrix-variate CWMs, for $G\in\left\{1,2,3\right\}$, are

\[\vecB_g= 
\begin{pmatrix*}[r]
    8.00 & 0.50 & 1.00 & 1.50\\
    1.00 & 1.00 & 0.50 & 1.50\\   
    4.00 & 1.00 & 1.00 & 1.50 \end{pmatrix*}, \quad
\vecA_{\vecX|g}=\vecA_{\vecY|g}=
\begin{pmatrix*}[r]
    1.50 & 1.00 & 1.00 & 1.00 \\
    1.50 & 1.00 &-1.00 &-1.00 \\    
   -1.00 & 1.00 & 1.50 & 1.50 \end{pmatrix*}, 
\]
\[\matsig_{\vecX|g}=\matsig_{\vecY|g}= 
\begin{pmatrix*}[r]
    1.00 & 0.80 & 0.64 \\
    0.80 & 1.00 & 0.80 \\    
    0.64 & 0.80 & 1.00 \end{pmatrix*}, \quad
\matPsi_{\vecX|g}=\matPsi_{\vecY|g}= 
\begin{pmatrix*}[r]
    1.50 & 0.90 & 0.54 & 0.32\\
    0.90 & 1.50 & 0.90 & 0.54\\    
    0.54 & 0.90 & 1.50 & 0.90\\
		0.32 & 0.54 & 0.90 & 1.50\end{pmatrix*}.
\]
For the covariate location, we take
\[\matm_{\vecX|1}= 
\begin{pmatrix*}[r]
    2.00 & 0.00 & 1.00 & 2.00 \\
    4.00 & 2.00 & 2.00 & 3.00 \\    
   -1.00 &-1.00 &-2.00 &-1.00 \end{pmatrix*},
\]
The other location matrices are obtained by adding a constant $c$ to each element of $\matm_{\vecX|1}$.
Specifically, we set $c$ equal to -3 and +3 for $\matm_{\vecX|2}$ and $\matm_{\vecX|3}$ under the ``close'' scenarios, respectively, whereas $c$  is set equal to -10 and +10 for $\matm_{\vecX|2}$ and $\matm_{\vecX|3}$ under the ``far'' scenarios, respectively.
The additional parameters, specific for each model and for $G\in\left\{1,2,3\right\}$ are: $\gamma_{\vecX|g}=\gamma_{\vecY|g}=7.00$ for the MVVG-MVVG CWM, $\omega_{\vecX|g}=3.00$, $\lambda_{\vecX|g}=-0.50$, $\nu_{\vecY|g}=10.00$ for the MVGH-MVST CWM, $\kappa_{\vecX|g}=1.20$ for the MVNIG-MVN CWM and $\omega_{\vecY|g}=3.00$, $\lambda_{\vecY|g}=-0.50$ for the MVN-MVGH CWM.

For each of the four matrix-variate CWMs, and each scenario, 100 datasets are generated and the corresponding model is fitted with $G=3$.
The MSEs of the regression coefficient estimates, computed over the 100 datasets, are reported in~\Cref{tab:res_par1,tab:res_par2,tab:res_par3,tab:res_par4}.
\begin{table}[!ht]
    \centering
\caption{MSE of the regression coefficient estimates over 100 datasets, for each scenario, for the MVVG-MVVG CWM.}
\label{tab:res_par1}       
\resizebox{\linewidth}{!}{
       \begin{tabular}{lccc}
				\hline\noalign{\smallskip}
Model & \multicolumn{3}{c}{MSE}  \\
MVVG-MVVG CWM   & Group 1 & Group 2 & Group 3 \\
	      \hline\noalign{\smallskip}
$N=200$ - close  & $\begin{pmatrix}  0.330 & 0.008 & 0.014 & 0.011\\
																         0.386 & 0.078 & 0.019 & 0.013 \\
																         0.446 & 0.006 & 0.018 & 0.013 \end{pmatrix}$ & $\begin{pmatrix} 0.357 & 0.005 & 0.016 & 0.010 \\
																																								                              0.400 & 0.006 & 0.019 & 0.012 \\
																																								                              0.490 & 0.007 & 0.025 & 0.012 \end{pmatrix}$ & $\begin{pmatrix}  0.410 & 0.006 & 0.019 & 0.011 \\
              0.378 & 0.006 & 0.018 & 0.011 \\
              0.419 & 0.008 & 0.016 & 0.011 \end{pmatrix}$ \\ [+14mm]
$N=500$ - close  & $\begin{pmatrix}  0.191 & 0.003 & 0.008 & 0.005\\
																         0.120 & 0.003 & 0.006 & 0.003 \\
																         0.179 & 0.003 & 0.007 & 0.005 \end{pmatrix}$ & $\begin{pmatrix} 0.171 & 0.003 & 0.007 & 0.004 \\
																																								                              0.121 & 0.002 & 0.005 & 0.004 \\
																																								                              0.130 & 0.002 & 0.006 & 0.004 \end{pmatrix}$ & $\begin{pmatrix}  0.147 & 0.003 & 0.006 & 0.003 \\
              0.173 & 0.002 & 0.006 & 0.004 \\
              0.158 & 0.002 & 0.007 & 0.004 \end{pmatrix}$ \\ [+14mm]
$N=200$ - far  & $\begin{pmatrix}  0.309 & 0.004 & 0.011 & 0.009\\
																       0.295 & 0.005 & 0.013 & 0.009 \\
																       0.263 & 0.006 & 0.013 & 0.007 \end{pmatrix}$ & $\begin{pmatrix} 0.690 & 0.006 & 0.015 & 0.010 \\
																																								                            0.707 & 0.005 & 0.015 & 0.009 \\
																																								                            0.623 & 0.005 & 0.019 & 0.009 \end{pmatrix}$ & $\begin{pmatrix}  0.750 & 0.004 & 0.013 & 0.011 \\
              0.950 & 0.005 & 0.021 & 0.012 \\
              0.879 & 0.006 & 0.020 & 0.013 \end{pmatrix}$ \\ [+14mm]
$N=500$ - far  & $\begin{pmatrix}  0.104 & 0.001 & 0.005 & 0.003\\
																       0.129 & 0.001 & 0.005 & 0.004 \\
																       0.111 & 0.002 & 0.005 & 0.003 \end{pmatrix}$ & $\begin{pmatrix} 0.235 & 0.002 & 0.005 & 0.003 \\
																																								                            0.223 & 0.002 & 0.006 & 0.004 \\
																																								                            0.229 & 0.002 & 0.005 & 0.003 \end{pmatrix}$ & $\begin{pmatrix}  0.316 & 0.002 & 0.004 & 0.003 \\
              0.277 & 0.003 & 0.006 & 0.004 \\
              0.216 & 0.002 & 0.005 & 0.003 \end{pmatrix}$ \\ 
\noalign{\smallskip}\hline
        \end{tabular}
}
\end{table}
\begin{table}[!ht]
    \centering
\caption{MSE of the regression coefficients estimates over 100 datasets, for each scenario, for the MVGH-MVST CWM.}
\label{tab:res_par2}       
\resizebox{\linewidth}{!}{
       \begin{tabular}{lccc}
				\hline\noalign{\smallskip}
Model & \multicolumn{3}{c}{MSE}  \\
MVGH-MVST CWM   & Group 1 & Group 2 & Group 3 \\
	      \hline\noalign{\smallskip}
$N=200$ - close  & $\begin{pmatrix}  0.419 & 0.007 & 0.020 & 0.013\\
																         0.400 & 0.007 & 0.022 & 0.018 \\
																         0.476 & 0.006 & 0.021 & 0.016 \end{pmatrix}$ & $\begin{pmatrix} 0.372 & 0.007 & 0.019 & 0.013 \\
																																								                              0.381 & 0.005 & 0.021 & 0.013 \\
																																								                              0.408 & 0.007 & 0.019 & 0.014 \end{pmatrix}$ & $\begin{pmatrix}  0.261 & 0.005 & 0.013 & 0.011 \\
              0.212 & 0.004 & 0.012 & 0.010 \\
              0.301 & 0.004 & 0.011 & 0.011 \end{pmatrix}$ \\ [+14mm]
$N=500$ - close  & $\begin{pmatrix}  0.119 & 0.003 & 0.008 & 0.005\\
																         0.122 & 0.002 & 0.007 & 0.006 \\
																         0.123 & 0.002 & 0.007 & 0.006 \end{pmatrix}$ & $\begin{pmatrix} 0.110 & 0.002 & 0.006 & 0.005 \\
																																								                              0.106 & 0.002 & 0.006 & 0.004 \\
																																								                              0.114 & 0.003 & 0.006 & 0.006 \end{pmatrix}$ & $\begin{pmatrix}  0.119 & 0.002 & 0.005 & 0.004 \\
              0.132 & 0.002 & 0.006 & 0.005 \\
              0.201 & 0.002 & 0.008 & 0.005 \end{pmatrix}$ \\ [+14mm]
$N=200$ - far  & $\begin{pmatrix}  0.207 & 0.003 & 0.010 & 0.008\\
																       0.344 & 0.004 & 0.018 & 0.010 \\
																       0.254 & 0.004 & 0.012 & 0.008 \end{pmatrix}$ & $\begin{pmatrix} 0.482 & 0.004 & 0.010 & 0.007 \\
																																								                            0.522 & 0.004 & 0.012 & 0.008 \\
																																								                            0.630 & 0.004 & 0.014 & 0.009 \end{pmatrix}$ & $\begin{pmatrix}  0.718 & 0.004 & 0.011 & 0.010 \\
              0.772 & 0.004 & 0.013 & 0.010 \\
              0.620 & 0.004 & 0.011 & 0.010 \end{pmatrix}$ \\ [+14mm]
$N=500$ - far  & $\begin{pmatrix}  0.096 & 0.001 & 0.004 & 0.003\\
																       0.103 & 0.002 & 0.005 & 0.003 \\
																       0.123 & 0.002 & 0.006 & 0.003 \end{pmatrix}$ & $\begin{pmatrix} 0.210 & 0.001 & 0.004 & 0.004 \\
																																								                            0.253 & 0.001 & 0.005 & 0.004 \\
																																								                            0.254 & 0.002 & 0.005 & 0.004 \end{pmatrix}$ & $\begin{pmatrix}  0.294 & 0.001 & 0.006 & 0.004 \\
              0.203 & 0.001 & 0.004 & 0.004 \\
              0.258 & 0.002 & 0.006 & 0.004 \end{pmatrix}$ \\ 
\noalign{\smallskip}\hline
        \end{tabular}
}
\end{table}
\begin{table}[!ht]
    \centering
\caption{MSE of the regression coefficients estimates over 100 datasets, for each scenario, for the MVNIG-MVN CWM.}
\label{tab:res_par3}       
\resizebox{\linewidth}{!}{
       \begin{tabular}{lccc}
				\hline\noalign{\smallskip}
Model & \multicolumn{3}{c}{MSE}  \\
MVNIG-MVN CWM   & Group 1 & Group 2 & Group 3 \\
	      \hline\noalign{\smallskip}
$N=200$ - close  & $\begin{pmatrix}  0.095 & 0.003 & 0.008 & 0.007\\
																         0.137 & 0.003 & 0.010 & 0.007 \\
																         0.149 & 0.003 & 0.011 & 0.009 \end{pmatrix}$ & $\begin{pmatrix} 0.235 & 0.004 & 0.012 & 0.007 \\
																																								                              0.217 & 0.004 & 0.012 & 0.007 \\
																																								                              0.255 & 0.004 & 0.016 & 0.010 \end{pmatrix}$ & $\begin{pmatrix}  0.162 & 0.002 & 0.008 & 0.008 \\
              0.185 & 0.002 & 0.008 & 0.007 \\
              0.177 & 0.002 & 0.008 & 0.007 \end{pmatrix}$ \\ [+14mm]
$N=500$ - close  & $\begin{pmatrix}  0.052 & 0.001 & 0.004 & 0.005\\
																         0.063 & 0.001 & 0.005 & 0.005 \\
																         0.043 & 0.001 & 0.004 & 0.004 \end{pmatrix}$ & $\begin{pmatrix} 0.090 & 0.002 & 0.007 & 0.004 \\
																																								                              0.097 & 0.021 & 0.008 & 0.004 \\
																																								                              0.100 & 0.002 & 0.007 & 0.005 \end{pmatrix}$ & $\begin{pmatrix}  0.061 & 0.001 & 0.003 & 0.003 \\
              0.065 & 0.001 & 0.003 & 0.003 \\
              0.063 & 0.001 & 0.003 & 0.003 \end{pmatrix}$ \\ [+14mm]
$N=200$ - far  & $\begin{pmatrix}  0.104 & 0.002 & 0.008 & 0.008\\
																       0.115 & 0.003 & 0.010 & 0.008 \\
																       0.125 & 0.003 & 0.010 & 0.008 \end{pmatrix}$ & $\begin{pmatrix} 0.480 & 0.002 & 0.009 & 0.008 \\
																																								                            0.518 & 0.003 & 0.010 & 0.008 \\
																																								                            0.481 & 0.003 & 0.115 & 0.007 \end{pmatrix}$ & $\begin{pmatrix}  0.445 & 0.002 & 0.008 & 0.008 \\
              0.490 & 0.002 & 0.009 & 0.008 \\
              0.538 & 0.002 & 0.010 & 0.007 \end{pmatrix}$ \\ [+14mm]
$N=500$ - far  & $\begin{pmatrix}  0.055 & 0.001 & 0.004 & 0.003\\
																       0.054 & 0.001 & 0.004 & 0.003 \\
																       0.043 & 0.001 & 0.003 & 0.003 \end{pmatrix}$ & $\begin{pmatrix} 0.184 & 0.001 & 0.004 & 0.003 \\
																																								                            0.155 & 0.008 & 0.004 & 0.003 \\
																																								                            0.188 & 0.001 & 0.004 & 0.003 \end{pmatrix}$ & $\begin{pmatrix}  0.229 & 0.001 & 0.004 & 0.004 \\
              0.212 & 0.001 & 0.004 & 0.003 \\
              0.228 & 0.001 & 0.004 & 0.003 \end{pmatrix}$ \\ 
\noalign{\smallskip}\hline
        \end{tabular}
}
\end{table}
\begin{table}[!ht]
    \centering
\caption{MSE of the regression coefficients estimates over 100 datasets, for each scenario, for the MVN-MVGH CWM.}
\label{tab:res_par4}       
\resizebox{\linewidth}{!}{
       \begin{tabular}{lccc}
				\hline\noalign{\smallskip}
Model & \multicolumn{3}{c}{MSE}  \\
MVN-MVGH CWM   & Group 1 & Group 2 & Group 3 \\
	      \hline\noalign{\smallskip}
$N=200$ - close  & $\begin{pmatrix}  0.326 & 0.009 & 0.018 & 0.010\\
																         0.235 & 0.009 & 0.017 & 0.009 \\
																         0.285 & 0.012 & 0.019 & 0.009 \end{pmatrix}$ & $\begin{pmatrix} 0.283 & 0.012 & 0.019 & 0.010 \\
																																								                              0.238 & 0.011 & 0.013 & 0.009 \\
																																								                              0.278 & 0.012 & 0.015 & 0.009 \end{pmatrix}$ & $\begin{pmatrix}  0.234 & 0.009 & 0.014 & 0.010 \\
              0.212 & 0.008 & 0.013 & 0.007 \\
              0.202 & 0.009 & 0.132 & 0.009 \end{pmatrix}$ \\ [+14mm]
$N=500$ - close  & $\begin{pmatrix}  0.064 & 0.004 & 0.005 & 0.003\\
																         0.078 & 0.004 & 0.007 & 0.003 \\
																         0.086 & 0.005 & 0.007 & 0.004 \end{pmatrix}$ & $\begin{pmatrix} 0.101 & 0.004 & 0.006 & 0.003 \\
																																								                              0.089 & 0.005 & 0.007 & 0.003 \\
																																								                              0.108 & 0.004 & 0.007 & 0.004 \end{pmatrix}$ & $\begin{pmatrix}  0.079 & 0.004 & 0.005 & 0.003 \\
              0.079 & 0.004 & 0.004 & 0.003 \\
              0.080 & 0.003 & 0.046 & 0.004 \end{pmatrix}$ \\ [+14mm]
$N=200$ - far  & $\begin{pmatrix}  0.156 & 0.007 & 0.012 & 0.007\\
																       0.147 & 0.008 & 0.012 & 0.006 \\
																       0.187 & 0.008 & 0.014 & 0.008 \end{pmatrix}$ & $\begin{pmatrix} 0.520 & 0.008 & 0.012 & 0.008 \\
																																								                            0.608 & 0.008 & 0.131 & 0.008 \\
																																								                            0.585 & 0.008 & 0.014 & 0.008 \end{pmatrix}$ & $\begin{pmatrix}  0.838 & 0.009 & 0.014 & 0.007 \\
              0.579 & 0.006 & 0.011 & 0.006 \\
              0.780 & 0.007 & 0.013 & 0.006 \end{pmatrix}$ \\ [+14mm]
$N=500$ - far  & $\begin{pmatrix}  0.068 & 0.003 & 0.006 & 0.003\\
																       0.062 & 0.003 & 0.005 & 0.003 \\
																       0.087 & 0.003 & 0.005 & 0.003 \end{pmatrix}$ & $\begin{pmatrix} 0.248 & 0.004 & 0.006 & 0.003 \\
																																								                            0.249 & 0.003 & 0.004 & 0.003 \\
																																								                            0.269 & 0.003 & 0.005 & 0.004 \end{pmatrix}$ & $\begin{pmatrix}  0.229 & 0.004 & 0.007 & 0.003 \\
              0.246 & 0.003 & 0.006 & 0.003 \\
              0.196 & 0.003 & 0.005 & 0.003 \end{pmatrix}$ \\ 
\noalign{\smallskip}\hline
        \end{tabular}
}
\end{table}

It is important to underline the well-known label switching issue, caused by the invariance of the likelihood function under relabeling the components of a mixture model \citep{fruhwirth06}. 
There are no generally accepted labeling methods, and considering the parameter set chosen, we simply attribute the labels by looking at the estimated $\matm_{\vecX|g}$.
In all cases considered the MSEs are quite small, meaning that the estimated parameters are close to their true values.
It should be noted that the MSEs for the intercepts are slightly larger, and the values are similar for all three groups when looking at the ``close" cases.
Hoverer, when the ``far'' cases are considered the intercepts of groups 2 and 3 produce higher MSEs compared to group 1.
This might depend on how far the groups are in the $\vecX$- and $\vecY$-spaces, so that small differences in the estimated slopes can produce big differences in the
estimates of the intercepts \citep{punzo15b}. 
Overall, the MSEs are slightly higher when the groups are close as compared to when the groups are well separated.
Finally, the MSEs improve with the increase of $N$, independent of the level of separation.

\subsection{Classification evaluation}
\label{sec:cc}

In this study, we first generate data from the MVN-MVN CWM, and then we apply the following transformation to the data: $\vecZ+\exp(\epsilon\vecZ)$, where $\vecZ\in\left\{\vecY,\vecX\right\}$ and $\epsilon>0$.
In this way, we are able to introduce right skewness which is governed by $\epsilon$.
Specifically, higher values of $\epsilon$ result in a higher level of skewness.
The parameters $\vecB_g,\matsig_{\vecX|g},\matsig_{\vecY|g},\matPsi_{\vecX|g},\matPsi_{\vecY|g}$, $g = 1,\ldots, G$, and $\matm_{\vecX|1}$ are the same as in Section~\ref{sec:pr}.
We set $N=200$ and, similarly to Section~\ref{sec:pr}, we add a constant $c$ to each element of $\matm_{\vecX|1}$ in order to obtain $\matm_{\vecX|2}$ and $\matm_{\vecX|3}$, respectively.
Specifically, we set $c=-30$ for $\matm_{\vecX|2}$ and $c=30$ for $\matm_{\vecX|3}$.
We consider $\epsilon\in\left\{0.6,1.0\right\}$, and for each pair ($\epsilon$, $N$) we generate 30 datasets.
For each dataset all the novel 24 matrix-variate CWMs and the MVN-MVN CWM are fitted for $G\in\left\{1,2,3,4\right\}$.
For each matrix-variate CWM, \tablename~\ref{tab:ari} reports the number of times the true $G$ is selected by the BIC, as well as the average ARI ($\overline{\text{ARI}}$) computed by considering the best fitting models over the 30 datasets.
\begin{table}[!ht]
    \centering
\caption{Number of times the true $G$ is selected by the BIC for each matrix-variate CWM, along with the average $\overline{\text{ARI}}$ computed by considering the best fitting models over the 30 datasets, when $\epsilon=0.6$ and $\epsilon=1.0$.}
\label{tab:ari}       
\resizebox{0.8\textwidth}{!}{
       \begin{tabular}{l|ccccc|ccccc}
				\hline
         & \multicolumn{5}{c}{$\epsilon=0.6$} & \multicolumn{5}{c}{$\epsilon=1.0$} \\
CWM      & G=1 & G=2 & G=3 & G=4 & $\overline{\text{ARI}}$ & G=1 & G=2 & G=3 & G=4 & $\overline{\text{ARI}}$ \\
	      \hline
MVST-MVST    & 0 & 2 & 28 & 0 & 0.97 & 0 & 0 & 27 & 3 & 0.93 \\
MVST-MVVG    & 0 & 2 & 28 & 0 & 0.97 & 0 & 0 & 27 & 3 & 0.92 \\
MVST-MVGH    & 0 & 2 & 28 & 0 & 0.97 & 0 & 0 & 28 & 2 & 0.91 \\
MVST-MVNIG   & 0 & 2 & 28 & 0 & 0.97 & 1 & 1 & 26 & 2 & 0.92 \\
MVST-MVN     & 0 & 2 & 28 & 0 & 0.97 & 0 & 0 & 20 & 10 & 0.79 \\
MVVG-MVST    & 0 & 2 & 28 & 0 & 0.97 & 0 & 2 & 27 & 1 & 0.89 \\
MVVG-MVVG    & 0 & 2 & 28 & 0 & 0.97 & 0 & 4 & 22 & 4 & 0.81 \\
MVVG-MVGH    & 0 & 2 & 28 & 0 & 0.97 & 1 & 1 & 27 & 1 & 0.83 \\
MVVG-MVNIG   & 0 & 2 & 28 & 0 & 0.97 & 1 & 3 & 20 & 6 & 0.81 \\
MVVG-MVN     & 0 & 2 & 28 & 0 & 0.97 & 0 & 1 & 18 & 11 & 0.54 \\
MVGH-MVST    & 0 & 2 & 28 & 0 & 0.97 & 0 & 0 & 27 & 3 & 0.96 \\
MVGH-MVVG    & 0 & 2 & 28 & 0 & 0.97 & 0 & 1 & 28 & 1 & 0.92 \\
MVGH-MVGH    & 0 & 2 & 28 & 0 & 0.97 & 0 & 1 & 28 & 1 & 0.93 \\
MVGH-MVNIG   & 0 & 1 & 29 & 0 & 0.98 & 0 & 1 & 28 & 1 & 0.96 \\
MVGH-MVN     & 0 & 2 & 28 & 0 & 0.97 & 0 & 0 & 21 & 9 & 0.69 \\
MVNIG-MVST   & 0 & 2 & 28 & 0 & 0.97 & 0 & 2 & 27 & 1 & 0.86 \\
MVNIG-MVVG   & 0 & 2 & 28 & 0 & 0.97 & 0 & 2 & 26 & 2 & 0.86 \\
MVNIG-MVGH   & 0 & 2 & 28 & 0 & 0.97 & 0 & 2 & 27 & 1 & 0.87 \\
MVNIG-MVNIG  & 0 & 2 & 28 & 0 & 0.97 & 0 & 1 & 28 & 1 & 0.87 \\
MVNIG-MVN    & 0 & 2 & 28 & 0 & 0.97 & 0 & 0 & 26 & 4 & 0.72 \\
MVN-MVST     & 0 & 1 & 28 & 1 & 0.98 & 0 & 2 & 11 & 17 & 0.75 \\
MVN-MVVG     & 0 & 1 & 28 & 1 & 0.98 & 0 & 1 & 11 & 18 & 0.73 \\
MVN-MVGH     & 0 & 1 & 28 & 1 & 0.98 & 0 & 2 & 10 & 18 & 0.71 \\
MVN-MVNIG    & 0 & 2 & 28 & 0 & 0.97 & 0 & 3 & 10 & 17 & 0.78 \\
MVN-MVN      & 0 & 1 & 26 & 3 & 0.98 & 0 & 0 & 11 & 19 & 0.47 \\
\noalign{\smallskip}\hline
        \end{tabular}
}
\end{table}

When $\epsilon=0.6$, the correct $G$ is practically always selected by all the matrix-variate CWMs, leading to nearly perfect classifications.
Interestingly, the matrix-variate CWMs having the MVN distribution for $f(\vecX;\bvtheta_{\vecX|g})$ produce slightly better $\overline{\text{ARI}}$ than the others.
However, when $\lambda=1$, the differences among the models appear evident because of the greater skewness.
Specifically, all the matrix-variate CWMs for which the MVN distribution is used in either $f(\vecX;\bvtheta_{\vecX|g})$ or $f(\vecY|\vecX;\bvtheta_{\vecY|g})$ show an overfitting tendency that leads to the selection of $G=4$ components most of the times.
Furthermore, it is interesting to note that this issue has a different magnitude depending on which one of $f(\vecX;\bvtheta_{\vecX|g})$ or $f(\vecY|\vecX;\bvtheta_{\vecY|g})$ is modelled using the MVN distribution.
In detail, such issue seems more relevant for the matrix-variate CWMs that use the MVN distribution for $f(\vecX;\bvtheta_{\vecX|g})$.
This has clear implications also on the resulting $\overline{\text{ARI}}$, that for these models assume lower values than the others.
As it is reasonable to expect, the MVN-MVN CWM produces the worst data classification ($\overline{\text{ARI}}=0.47$).

For the matrix-variate CWMs assuming a skewed distribution both in $f(\vecX;\bvtheta_{\vecX|g})$ and $f(\vecY|\vecX;\bvtheta_{\vecY|g})$, the true $G$ is almost always chosen, with only the MVVG-MVNIG CWM showing similar performance to the matrix-variate CWMs having the MVN distribution for $f(\vecY|\vecX;\bvtheta_{\vecY|g})$.
In terms of classification, the $\overline{\text{ARI}}$ assume quite high values, despite they are slightly lower than the $\epsilon=0.6$ case.
   
\section{Real Data Analyses}
\label{sec:real}

\subsection{Overview}

In this section, all the matrix-variate CWMs discussed so far are applied to two real datasets.
For comparison purposes, the matrix-variate FMRs based on the skewed distributions analyzed in this manuscript, as well as the MVN FMR, are also considered.

\subsection{Data}
\label{sec:data}

The first application concerns a dataset referred to as ``Education'', provided by the Italian national agency for the evaluation of universities and research institutes.
It contains $p=2$ indicators measuring the satisfaction of the students which we use as response variables: (i) the percentage of graduates who would enroll again in the same degree program and (ii) the percentage of students satisfied with the degree program. 
The data also includes $q=2$ indicators related to the academic careers of the students and the results of the training activities which we use as covariates: (i) the percentage of students that have earned at least 40 course credits during the solar year and (ii) the percentage of students who continue to the second year in the same degree program having acquired at least 20 credits in the first year.
Both sets of variables are measured over $r=3$ years for $N=74$ degree programs in the universities of southern Italy.
There are two groups of degree programs in the data, namely $N_1=32$ bachelor's degrees and $N_2=42$ master's degrees.

The second application considers the \texttt{Insurance} dataset included in the \textbf{splm} package \citep{splm}.
This dataset was introduced by \citet{millo2011non} to study the consumption of non-life insurance across the $N=103$ Italian provinces for the years 1998–2002, an it has been also recently considered in \citet{tomarchio2021} for the MVN-MVN CWM.
As was done by \citet{tomarchio2021}, we consider $p=2$ variables related to the consumption and the presence of insurance products in the market as response variables: (i) the real per-capita non-life premiums in 2,000 euros and (ii) the density of insurance agencies per 1,000 inhabitants used 
For the covariates, we consider $q=3$ variables: (i) the real per-capita GDP, (ii) the real per-capita bank deposits and (iii) the real interest rate on lending to families and small enterprises. 
These are regularly used in the literature as proxies for general level of economic activity, stock of wealth, and opportunity cost of allocate funds in insurance policies, respectively.
Unlike the first application, we do not have a ``ground truth'' classification of the data, and therefore we cannot compute the ARI to evaluate the partitions of the competing models.
However, the findings of \citet{millo2011non} and \citet{tomarchio2021} underline the existence of two macro areas, namely Central-Northern Italy, characterized by an insurance penetration level relatively close to the European averages, and Southern Italy, where a general economic underdevelopment has long been standing as a fundamental social and political problem.
A graphical analysis can be useful for assessing the quality of the classification produced.

\subsection{Results}
\label{sec:res}

In both applications, all the matrix-variate CWMs and FMRs are fitted for $G\in\left\{1,2,3\right\}$.
When the Education dataset is considered, the best matrix-variate CWM, according to the BIC, is the MVN-MVVG model with $G=2$, whereas the best matrix-variate FMR is the MVST with $G=1$.
The classification results give an ARI of 0.84 for the MVGH-MVST CWM, i.e., a very good classification, and an ARI of 0 for the MVST FMR.
Such a behavior of FMR models is not uncommon, as shown for example in \citet{tomarchio2021} for the MVN FMR, where only one group was detected in their data.
Here, such a problem is present in our data even if skewed matrix-variate distributions are used.

\begin{figure}[!ht]
\centering
\subfigure[]
{\includegraphics[width=0.324\textwidth]{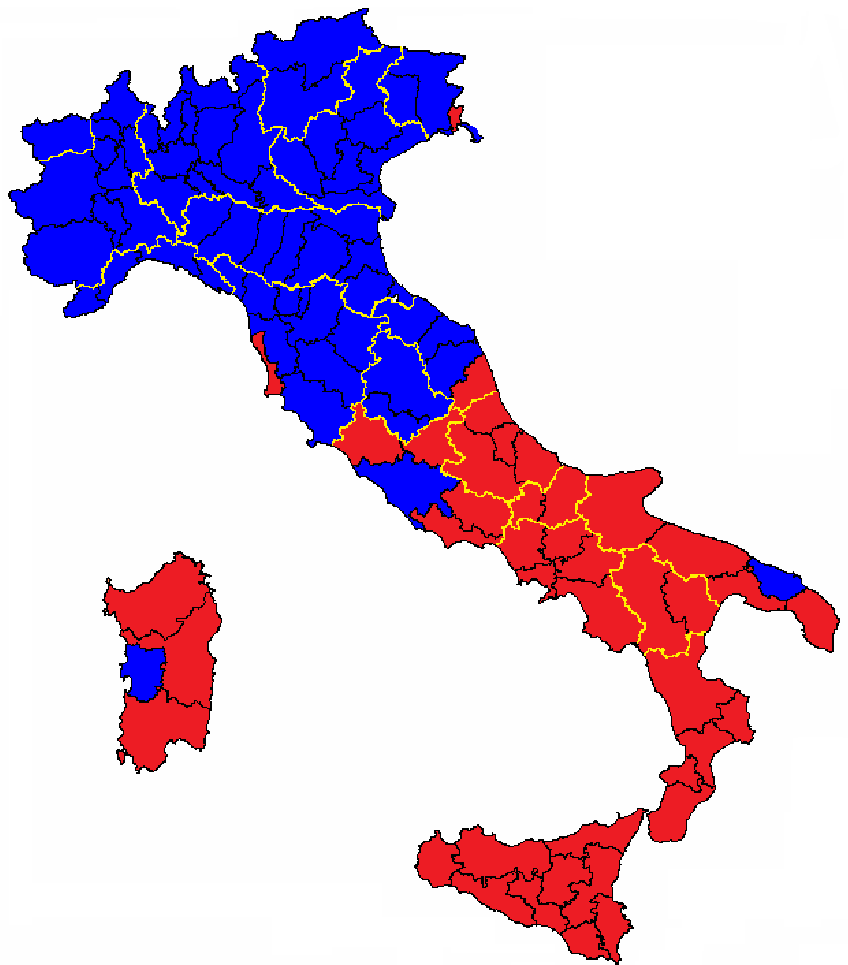}}
\subfigure[]
{\includegraphics[width=0.324\textwidth]{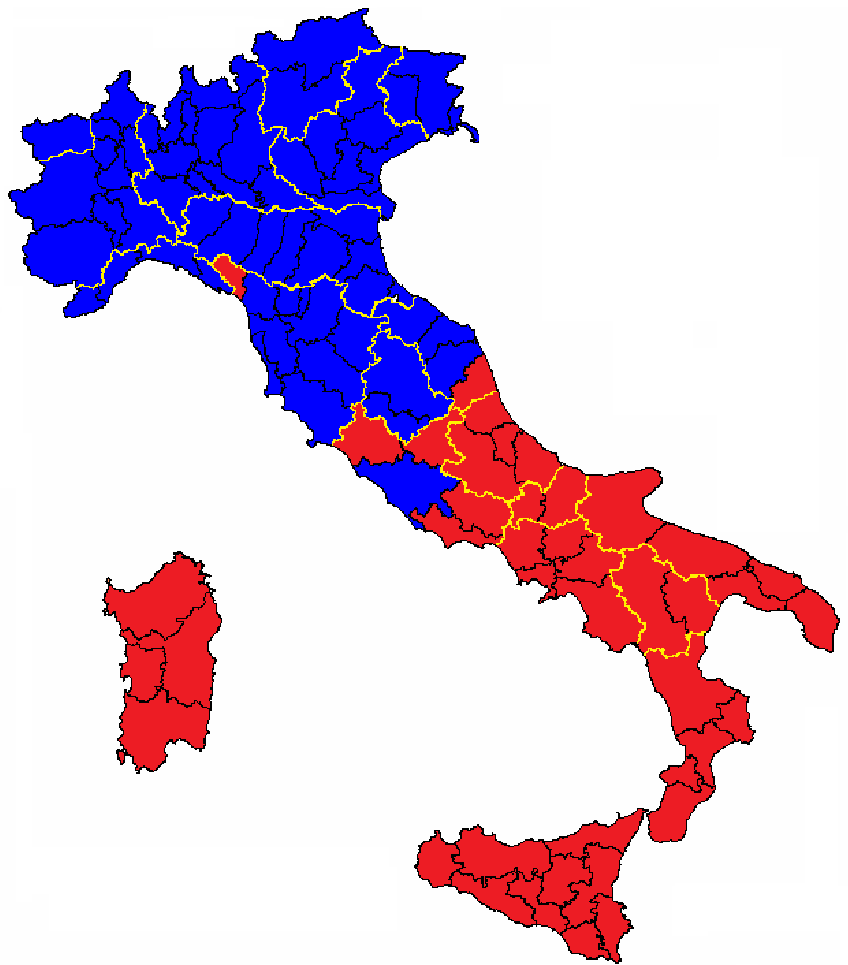}}
\subfigure[]
{\includegraphics[width=0.324\textwidth]{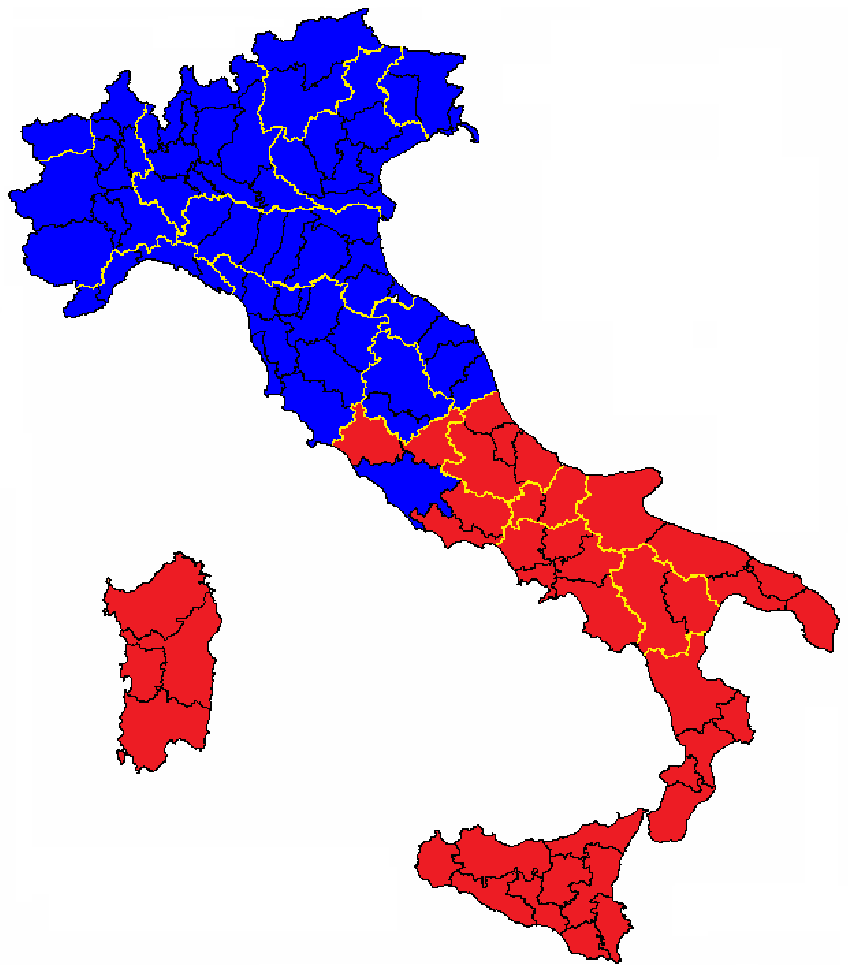}}
\caption{Partitions produced by the MVST FMR (a), MVN-MVN CWM (b) and MVVG-MVST CWM (c) for the insurance data.
}
\label{fig:col}
\end{figure}

For the \texttt{Insurance} dataset, the best matrix-variate CWM according to the BIC is the MVVG-MVST with $G=2$, whereas the best matrix-variate FMR is the MVST with $G=2$.
Therefore, in this application, both approaches agree in detecting two groups in the data, unlike \citet{tomarchio2021} where the MVN FMR found $G=3$ groups in the data.
Additionally, the use of skewed matrix-variate distributions provide a better fit than the corresponding normal models.
These two partitions are illustrated in~\figurename~\ref{fig:col} using the Italian political map.
Specifically, the Italian regions are bordered in yellow (islands excluded), while the internal provinces are delimited with the black lines and colored according to the estimated group membership both for the MVVG-MVST CWM and the MVST FMR.
We also show the map for the MVN-MVN CWM illustrated in \citet{tomarchio2021} for comparison purposes.
A few interesting points are now discussed.
We notice that although the MVST-FMR roughly recognizes the Central-Northern Italy and the Southern Italy groups, these two groups put together some provinces that span all over the country without a straightforward and reasonable justification.
For the MVN-MVN CWM, with the exclusion of three cases, all the provinces belonging to the same region are clustered together.
These three exceptions concern the province of Rome (in the Lazio region), which due to its social-economic development is reasonably assigned to the Central-Northern Italy group, the province of Ascoli-Piceno (in the Marche region) and, in particular, the province of Massa-Carrara (in the Toscana region), which is unreasonably assigned to the Southern Italy group.
On the other hand, in addition to producing a higher BIC, the MVVG-MVST perfectly divides Italy in two macro areas, where all the provinces belonging to the same region are clustered together, with the reasonable exception of the province of Rome.

\section{Conclusions}
\label{sec:end}

A novel family of 24 matrix-variate CWMs was introduced.
Extending the matrix-variate normal CWM recently introduced by \cite{tomarchio2021}, the distributions of the responses and of the covariates were allowed to be skewed in each cluster.
We specifically considered the matrix-variate skew-$t$, generalized hyperbolic, variance-gamma and normal inverse Gaussian distributions.
In addition, by also considering the matrix-variate normal distribution, our models were flexible enough to handle scenarios in which the covariates and the responses conditioned on the covariates are skewed, or in which one of the two sets of variables is normally distributed and the other one is skewed.
As a by-product, the four skewed matrix-variate FRMs were also introduced.

An ECM algorithm was discussed for parameter estimation, and its capability of recovering the parameters of the data generating model was tested under several scenarios by means of a simulation study.
A comparison among the matrix-variate CWMs in terms of classification performance, as well as the capability of the BIC to detect the underlying group structure of the data, was also investigated via simulated data.
All of the 24 novel matrix-variate CWMs, as well as the matrix-variate normal CWM and matrix-variate FMRs, were fitted to two real datasets.
The results of the first application show that the matrix-variate FMRs, even using skewed matrix-variate distributions, might fail to properly model the data, whereas the best matrix-variate CWM properly identified the two groups.
Furthermore, the best BIC was achieved by one of the skewed matrix-variate CWM models over the matrix-variate normal CWM.

In the second application, although lacking a true classification, an underlying group structure is supported by the existing literature.
In such a case, one of our skewed matrix-variate CWMs provides a better fit and classification than the matrix-variate normal CWM and the best among the matrix-variate FMRs.

Further model developments can be readily proposed.
Specifically, constrained parameterizations of the covariance matrices can be employed, both for the distribution of the responses and the distribution of the covariates.
This can be done by following two different routes: (i) the eigen decomposition approach in the fashion of \citet{sarkar2020parsimonious} or (ii) the bilinear factor analyzers method in accordance to \citet{gallaugher2019mixtures}.
Both proposals can drastically reduce the number of estimated parameters, allowing for more parsimonious models.
\section*{Acknowledgements}
This work was funded by a Vanier Canada Graduate Scholarship and Banting Postdoctoral Fellowship (Gallaugher), the Canada Research Chairs Program and The Steacie Memorial Fund (McNicholas).

\bibliography{SkewCWM.bib}

\appendix
\section*{Appendix}

\section{Specific distribution ECM updates}
\label{sec:App1}

\subsubsection*{Matrix-Variate Skew-$t$ Distribution}
\label{sec:st}

In the case of the MVST distribution, we need to update $\nu_g$.
However, a closed-form expression is not available, and thus needs to be updated numerically.
When $\fX$ is considered, the update for $\nu_{\vecX|g}$ is obtained by solving the equation
\begin{equation}
\log\left(\frac{\nu_{\vecX|g}}{2}\right)+1-\varphi\left(\frac{\nu_{\vecX|g}}{2}\right)-\frac{1}{\ddot{T}_g}\sum_{i=1}^N\ddot{z}_{ig}(\ddot{m}_{ig\vecX}+\ddot{n}_{ig\vecX})=0,
\label{eq:nugup}
\end{equation}
where $\varphi(\cdot)$ denotes the digamma function.
Conversely, when $\fY|\vecX$ is considered, the update for $\nu_{\vecY|g}$ is obtained via~\eqref{eq:nugup}, but $\nu_{\vecX|g}$, $\ddot{m}_{ig\vecX}$ and $\ddot{n}_{ig\vecX}$ are replaced by $\nu_{\vecY|g}$, $\ddot{m}_{ig\vecY}$ and $\ddot{n}_{ig\vecY}$, respectively.

\subsubsection*{Matrix-Variate Generalized Hyperbolic Distribution}
\label{sec:gh}

For the MVGH distribution, we need to update $\lambda_g$ and $\omega_g$.
Also in this case there are not closed-form expressions.
The resulting updates, when $\vecX$ is considered, are
\begin{align}
\ddot{\lambda}_{\vecX|g}&=\overline{n}_{\vecX|g}\dot{\lambda}_{\vecX|g}\left[\left.\frac{\partial}{\partial s}\log(K_{s}(\dot{\omega}_{\vecX|g}))\right|_{s=\dot{\lambda}_{\vecX|g}}\right]^{-1} \label{eq:lamup}\\
\ddot{\omega}_{\vecX|g}&=\dot{\omega}_{\vecX|g}-\left[\left.\frac{\partial}{\partial s}q(\ddot{\lambda}_{\vecX|g},s)\right|_{s=\dot{\omega}_{\vecX|g}}\right]\left[\left.\frac{\partial^2}{\partial s^2}q(\ddot{\lambda}_{\vecX|g},s)\right|_{s=\dot{\omega}_{\vecX|g}}\right]^{-1},
\label{eq:omup}
\end{align}
where the derivative in \eqref{eq:lamup} is computed numerically, 
$$
q(\ddot{\lambda}_{\vecX|g},\dot{\omega}_{\vecX|g})=\sum_{i=1}^N \ddot{z}_{ig}\left[\log(K_{\ddot{\lambda}_{\vecX|g}}(\dot{\omega}_{\vecX|g}))-\ddot{\lambda}_{\vecX|g}\overline{n}_{\vecX|g}-\frac{1}{2}\dot{\omega}_{\vecX|g}\left(\overline{l}_{\vecX|g}+\overline{m}_{\vecX|g}\right)\right],
$$
and $\overline{n}_{\vecX|g}=({1}/{\ddot{T}_g})\sum_{i=1}^N \ddot{z}_{ig}\ddot{n}_{ig\vecX}$.
When $\fY|\vecX$ is considered, we need to replace $\lambda_{\vecX|g}$, $\omega_{\vecX|g}$, $\overline{l}_{\vecX|g}$, $\overline{m}_{\vecX|g}$, and $\overline{n}_{\vecX|g}$ with $\lambda_{\vecY|g}$, $\omega_{\vecY|g}$, $\overline{l}_{\vecY|g}$, $\overline{m}_{\vecY|g}$, and $\overline{n}_{\vecY|g}$, respectively, where $\overline{m}_{\vecY|g}=(1/\ddot{T}_g)\sum_{i=1}^N\ddot{z}_{ig}\ddot{m}_{ig\vecY}$ and $\overline{n}_{\vecY|g}=({1}/{\ddot{T}_g})\sum_{i=1}^N\ddot{z}_{ig}\ddot{n}_{ig\vecY}$.

\subsubsection*{Matrix-Variate Variance-Gamma Distribution}
\label{sec:vg}

Similarly to the previous distributions, there is not a closed-form expression to update $\gamma_g$ for the MVVG distribution. 
Thus, when $\fX$ is considered, this update is obtained by solving the equation
\begin{equation}
\log\gamma_{\vecX|g}+1-\varphi(\gamma_{\vecX|g})+\overline{n}_{\vecX|g}-\overline{l}_{\vecX|g}=0.
\label{eq:gammup}
\end{equation}
Conversly, when $\fY|\vecX$ is considered, we replace $\gamma_{\vecX|g}$, $\overline{n}_{\vecX|g}$ and $\overline{l}_{\vecX|g}$ with $\gamma_{\vecY|g}$, $\overline{n}_{\vecY|g}$ and $\overline{l}_{\vecY|g}$, respectively.

\subsubsection*{Matrix-Variate Normal Inverse Gaussian Distribution}
\label{sec:nig}

The MVNIG distribution is the only having a closed form expression for its additional parameter.
Specifically, when $\fX$ is considered, the update for $\kappa_{\vecX|g}$ is
$$
\ddot{\kappa}_{\vecX|g}=\frac{1}{\overline{l}_{\vecX|g}}.
$$
If $\fY|\vecX$ is considered, $\ddot{\kappa}_{\vecX|g}$ and $\overline{l}_{\vecX|g}$ are replaced with $\ddot{\kappa}_{\vecY|g}$ and $\overline{l}_{\vecY|g}$, respectively.

\end{document}